# Considerazioni sulla curva di luce della Nova Delphini 2013

Costantino Sigismondi ICRA, sigismondi@icra.it  XXI GAD meeting, La Spezia, 12 ottobre 2013

**Abstract** E' possibile valutare la magnitudine assoluta della Nova Delphini 2013 già con osservazioni visuali dell'AAVSO, seguendo i lavori di Leonida Rosino ad Asiago sulle novae di M31. Si fornisce un modello interpretativo naïf della curva di luce basato sulla legge esponenziale decrescente, allo scopo di rendere più partecipi del complicato quadro teorico generale anche gli astrofili.

## Introduzione

Scoperta il 14 agosto è arrivata nello spazio di un giorno alla quarta grandezza per poi iniziare la sua diminuzione. Oltre quattrocento astrofili in tutto il mondo registrati all'AAVSO (American Association of Variable Stars Observers) stanno seguendo l'evoluzione di questa "stella nuova" con metodi digitali e visuali. Rispetto al passato, grazie al web, la comunità degli astrofili puo' muoversi compatta, senza che si perda tempo prezioso, in caso di eventi rari di questo tipo. La circolare dell'AAVSO è arrivata ai destinatari della mailing list il 16 agosto, con la curva di luce ancora in fase ascendente, meno di 48 ore dopo la scoperta.

I più «scettici», come me, hanno avuto ancora 3 giorni di tempo per decidersi a puntare un binocolo verso la regione celeste contornata dal caratteristico rombo del Delfino e dalla "tromba" della Sagitta: di solito gli oggetti annunciati dall'AAVSO o sono troppo deboli per un binocolo urbano, o sono quasi immersi nella luce solare a latitudini australi da centro galattico, ma non era il caso di questa Nova. Meno male che è venuto un ospite a casa il 19 agosto, e ho pensato di offrirgli uno spettacolo insolito, che mi aspettavo avesse solo un sapore storico. Infatti una stella di quinta grandezza non impressiona quasi nessuno... a meno che non si comprenda l'importanza sia storica che astrofisica di uno dei fenomeni più affascinanti del cielo: le stelle novae.

Ed effettivamente, prima motivato dalla possibilità di poter contribuire al database dell'AAVSO anche con le mie osservazioni, e poi motivato dall'evento in sé, piuttosto raro come 25 per secolo, ho dedicato tutte le sere un tempo per le osservazioni come non facevo da tanto tempo, essendomi dedicato al Sole con un impegno quasi esclusivo. Ho fatto trenta notti, fino ad ora, su quella Nova.

## Novae e Supernovae ad Asiago

Presto mi è tornata alla mente S Andromedae, esplosa nell'agosto del 1885 quando ancora non si sapeva che M31 era una galassia, e che questa galassia era a ben 3 milioni di anni luce da noi.
Sono andato a rileggere le pagine di Piero Tempesti sull'Enciclopedia dell'Astronomia, immortale capolavoro, edito da A. Curcio nel 1983, trovando molte risposte.
Tempesti si rifaceva ai lavori svolti ad Asiago col telescopio Galileo da 122 cm al Pennar, quando ancora non era stato costruito il telescopio Copernico (1973) di 180 cm a Cima Ekar. Ho avuto l'opportunità di osservare per 2 settimane in ottobre 2012 e febbraio-marzo 2013 all'osservatorio di Asiago, standoci notte e di', ma solo ora che ho osservato anche io una Nova, sia pure solo visualmente, ho potuto apprezzare a pieno il lavoro svolto da quegli astronomi nei settanta anni di vita dell'osservatorio fondato nel 1942.

Effettivamente questa è stata per me, che sono fisico di formazione, l'occasione per apprezzare quanto la scuola di Asiago-Padova abbia svolto un servizio molto importante per l'astronomia delle Novae, segnatamente per quelle della galassia di Andromeda, che si trovano tutte alla medesima distanza da noi. Con i lavori di Leonida Rosino e collaboratori è stato possibile stabilire una relazione statistica tra luminosità assoluta e rapidità della fase di discesa, come esprime la seguente figura [ottenuta dalla tabella a p. 427 di Tempesti, 1983, presa da L. Rosino].

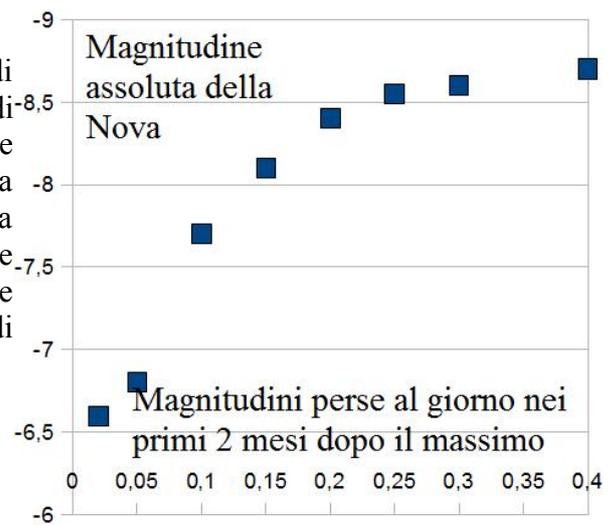

Lo stesso autore, che spiega anche la genesi, negli anni venti del '900, del nome Supernova, per distinguere dalle Novae che raggiungono luminosità nettamente inferiori [Tempesti, 1983, p. 444], ha anche trattato della frequenza statistica delle Supernovae, una per galassia ogni 50 anni, discutendo anche sulla prossima Supernova galattica, argomento che pure io trattai al congresso UAI di Piombino [Sigismondi, 2005].

In termini di nomenclatura vanno segnalate anche le più recenti **Hypernovae** [B. Paczynski, 1998], chiamate in causa anche per spiegare i Gamma Ray Burst, che in termini energetici sono le esplosioni più grandi dell'Universo. Oggetti a distanze cosmologiche, ben più lontani della Galassia di Andromeda, anche mille volte tanto, che raggiungono per qualche istante anche magnitudini visibili ad occhio nudo, come è accaduto in un caso: la sesta. Se fossero nella galassia di Andromeda, ad un Mega Parsec potrebbero splendere fino alla magnitudine -9, se fossero a 10 Parsec, avremmo ancora 12 magnitudini da togliere per avere la magnitudine assoluta, arrivando alla -21. Ma sulla opportunità di considerare le Hypernovae come una nuova classe non mi pronuncio, essendo ancora in pieno svolgimento il dibattito circa la natura dei Gamma Ray Bursts [es. Ruffini e Penacchioni, 2013; Pisani et al., 2013].

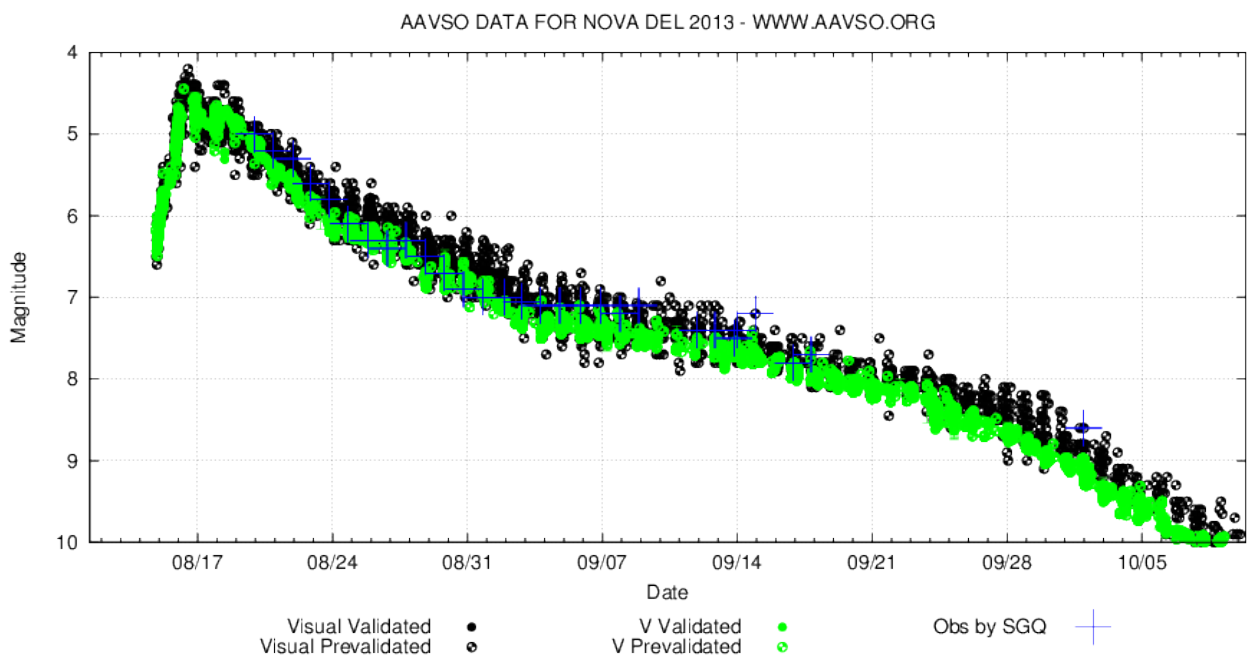

Tornando alla nostra nova nel Delfino, come in tanti altri casi analoghi, la curva di luce segue una legge esponenziale decrescente, messa in evidenza dal grafico magnitudine-tempo che segue un andamento discendente abbastanza rettilineo (vedi figura; le croci blu sono le mie osservazioni, fino al 1 ottobre; è molto interessante notare che anche con i soli dati miei si puo' calcolare la magnitudine assoluta della Nova come di seguito).
Il gioco tra aumento dell'area del gas in espansione e diminuzione della sua densità fa in modo che la pseudofotosfera prima si espanda e poi si ritragga. Qui entra in causa la teoria delle atmosfere stellari, che costituisce un capitolo fondamentale dell'astrofisica classica.

La nova ha raggiunto il picco il 16 agosto a 4.3 ed il 10 ottobre si attesta attorno alla magnitudine 9.8. In 55 giorni ha perso 5.5 magnitudini, dunque una discesa media di 0.1 magnitudini al giorno. Usando la tabella di Rosino, graficata nella figura 1, possiamo attribuire a questa Nova una magnitudine assoluta di -7.7. Possiamo fare l'esercizio di calcolare il modulo di distanza di

4.3+7.7=12 magnitudini. Se non ci fosse assorbimento interstellare (che nel piano della Galassia è massiccio) questa Nova dovrebbe trovarsi ad oltre un Megaparsec da noi, cioè ben fuori dalla Galassia, ma possiamo dire con certezza che questo indebolimento è dovuto principalmente alle polveri galattiche lungo la linea di vista.

**La curva di luce ed i fenomeni fisici descritti con una legge esponenziale descrescente**
In termini di energetica la radiazione che noi riceviamo nel nostro telescopio dalla Nova è una frazione dell'energia che essa disperde nello spazio in seguito all'esplosione che è occorsa nel sistema stellare che la costituisce. Ci si puo' chiedere come mai la luce segua sempre una legge esponenziale decrescente.
Innanzitutto va detto che questo tipo di legge matematica si ritrova in tantissimi fenomeni fisici.
La ragione puo' essere compresa considerando l'equazione differenziale che sta all'origine della legge esponenziale decrescente: $dI/dt = -cI$
Questa equazione ben compatta si "legge" come segue: la quantità di energia *dI* rilasciata (segno negativo) nell'intervallo di tempo *dt* è proporzionale secondo la costante *c* all'energia totale *I*, oppure, in altre parole ancora, è una percentuale dell'energia totale *I*.
Continuando a "leggere" questa equazione, risolvendola -per cosi' dire- con le parole, si puo' dire che quando all'inizio l'energia dell'esplosione *I* è grande, il ritmo *dI/dt* con cui questa si disperde è rapido; poi quando *I* diminuisce anche il ritmo della sua dispersione diminuisce. Questo significa che il processo rallenta sempre di più, dando luogo alla tipica discesa di un esponenziale decrescente, per il quale la diminuzione percentuale diminuisce al diminuire dell'energia residua.
Non solo le stelle novae ubbidiscono a questa legge, ma anche tutti i fenomeni di termalizzazione, e molti fenomeni di smorzamento o spegnimento, anche biologici, come ad esempio l'attività dei lieviti in funzione della temperatura ambiente...

**Conclusioni: una previsione per il ritorno nei ranghi della Nova Delphini 2013 su basi fisiche**
Ogni volta che ci troviamo in presenza di un esponenziale decrescente si puo' valutare un tempo caratteristico del fenomeno, detto *τ* **tau**, che è il tempo in capo al quale l'intensità è scesa di un fattore 1/e=0.36 della differenza tra il valore massimo e quello minimo. Si puo' dire che dopo cinque volte questo valore tau, la termalizzazione sia avvenuta, se si tratta di un fenomeno di scambi di calore, come ben sanno i ragazzi del biennio del tecnico industriale, che misurano la costante tau di un termometro nel laboratorio di scuola.
Nel caso della Nova Delphini 2013 noi sappiamo già che al minimo era una stella di magnitudine 16. Dunque il gap tra minimo e massimo è stato di 12 magnitudini. Il tempo tau in questo caso sarebbe quello necessario a che la stessa cali di 12/e~4.4 magnitudini.
Da quando la ho osservata di magnitudo 5 il 19 agosto 2013, proprio in questi primi giorni di ottobre 2013, ha superato la 9 e si avvia verso la decima. Dunque possiamo dire che il tau di questa nova è attorno ai 2 mesi, e ci possiamo aspettare che in 10-12 mesi la stella progenitrice ritorni alla sua condizione iniziale.
In realtà le Novae hanno un comportamento più complicato della semplice legge esponenziale decrescente, come afferma ancora Tempesti [1983, p. 429] «circa 3.5 magnitudini sotto il massimo la Nova entra nella fase nebulare» e lo spettro continuo cede il passo ad uno di righe di emissione.
Quindi il meccanismo di emissione passa dal continuo alle righe e la pendenza della curva di luce cambia. In generale le Novae possono impiegare alcuni anni a ritornare allo stadio iniziale.